# Popularity, face and voice: Predicting and interpreting livestreamers' retail performance using machine learning techniques

Xiong Xiong[1,3], Fan Yang[2], Li Su[3,*]

1. School of Information and Communication Engineering, Beijing University of Posts and Telecommunications, Beijing 100876, China
2. School of Economics and Management, Southeast University, Nanjing 211189, China
3. Department of Neuroscience, University of Sheffield, Sheffield S10 2TN, UK

   Corresponding author: Li Su

**Abstract:** Livestreaming commerce, a hybrid of e-commerce and self-media, has expanded the broad spectrum of traditional sales performance determinants. To investigate the factors that contribute to the success of livestreaming commerce, we construct a longitudinal firm-level database with 19,175 observations, covering an entire livestreaming subsector. By comparing the forecasting accuracy of eight machine learning models, we identify a random forest model that provides the best prediction of gross merchandise volume (GMV). Furthermore, we utilize explainable artificial intelligence to open the black-box of machine learning model, discovering four new facts: 1) variables representing the popularity of livestreaming events are crucial features in predicting GMV. And voice attributes are more important than appearance; 2) popularity is a major determinant of sales for female hosts, while vocal aesthetics is more decisive for their male counterparts; 3) merits and drawbacks of the voice are not equally valued in the livestreaming market; 4) based on changes of comments, page views and likes, sales growth can be divided into three stages. Finally, we innovatively propose a 3D-SHAP diagram that demonstrates the relationship between predicting feature importance, target variable, and its predictors. This diagram identifies bottlenecks for both beginner and top livestreamers, providing insights into ways to optimize their sales performance.

# 1. Introduction

From traditional physical distribution to online shopping and now to e-commerce live streaming, the retail economy has undergone three wave of reforms (Zhang et al., 2023). Livestreaming commerce is distinct from its predecessors as it carries the characteristics of both E-commerce and self-media (Aytan, 2021), potentially extending the spectrum of historical determinants of sales performance (Verbeke, 1997) (Chawla et al., 2020). Unlike traditional retailers, livestreamers can conduct parasocial interactions via digital platform, thus building relationships with numerous potential buyers simultaneously. In comparison to traditional e-commerce, the physical appearance and voice of broadcasters may evoke an aesthetic and intimate feeling among customers, which can enhance their customer loyalty and stimulate their willingness to purchase.

The nexus of livestreaming literature is Gross Merchandise Volume (GMV) and its determinants. By investigating the factors that contribute to GMV, researchers and practitioners can gain valuable insights into the underlying mechanisms of consumer behavior in the context of livestreaming e-commerce, as well as develop effective strategies to improve sales performance and predict the revenue of a live broadcast. Table 1 provides detailed information on the literature that focuses on livestreaming commerce.

**Table 1**

Literature review of e-commerce live streaming (ELS).

| Author | Topic | Main potential influencer | Data source | Observations | Methods |
|---|---|---|---|---|---|
| (Liu et al., 2022) | Investigates the effects of live-streaming interactivity, authenticity, and entertainment on purchase intention in the field of ELS | Interactivity, authenticity, and entertainment of ELS | Questionnaire | 357 | SOR |
| (Xu et al., 2021) | Investigates the ways in which live streaming impacts consumer purchasing intentions in the cross-border ELS | Affordance of live streaming, transparency and platform satisfaction | Questionnaire | 272 | SOR |
| (Dong et al., 2022) | Investigates the live-streaming factors that increase consumer purchasing in the field of ELS for green agricultural products | Information quality, system quality, service quality, telepresence, and social presence | Questionnaire | 726 | SEM |
| (Yang et al., 2022) | Investigate the pathways through which ELS induces impulse consumers' purchasing intention | Interface design, live atmosphere, visual appeal, perceived arousal and engagement | Questionnaire | 339 | SOR + PLS-SEM |
| (Guo et al., 2021) | Investigate the impulse buying behavior triggered by live streaming in cross-border e-commerce | Perceived cost savings | Questionnaire | 272 | SOR |
| (Yu et al., 2022) | Investigate the factors that influence consumers' engagement with streamers in the context of ELS | Enthusiasm and preparation of streamers | EEG of the subject while experiencing the ELS purchase | 43 | LSD post-hoc analysis |
| (Liu & Yu, 2022) | Investigate the role of streamers' personal charm and influence in promoting consumers' purchase intention | Streamers' personal charm, blogger's influence, commodity type | Recruit subjects to experience ELS | 398 | SOR |
| (Lin et al., 2022) | Investigate the stimulating factors of consumers' perceived enjoyment in ELS | Demand, convenience, interactivity, and playfulness | Questionnaire | 335 | SOR |
| (Qing & Jin, 2022) | Investigate the influence ways of service quality, information quality, and system quality on consumers' purchase intention | Service quality, information quality, and system quality | Questionnaire | 231 | Hypothesis testing |
| (Li et al., 2022) | Investigate the effect of social presence in ELS on customer impulse buying | Social presence | Questionnaire | 189 | SOR |
| (Dang-Van et al., 2023) | Investigate the influence of broadcasters' physical attractiveness on consumer engagement in ELS | Broadcasters' physical attractiveness | Questionnaire | 810 | SEM |
| (Jiang et al., 2022) | Investigate the factors underlying consumer engagement in the entrepreneurs' live streaming | Reputation, expertise, and interactivity, guarantee, authenticity, and money-saving | Questionnaire | 231 | SEM |

**Table 1**

Literature review of e-commerce live streaming (ELS) (continued).

| | | | | | |
|---|---|---|---|---|---|
| (Clement Addo et al., 2021) | Investigate the influence of social factors on customer participation in ELS | Likes, chats, visits and following time | Questionnaire | 1726 | Hypothesis testing |
| (X. Wang et al., 2022) | Investigate the impact on consumer attitudes in the context of Chinese online influencers' ELS | Expertise, bargaining power, post-sales services, and live streaming schedules | Questionnaire | 430 | SOR |
| (Ma et al., 2022) | Investigate the psychological mechanisms of how live peculiarities impact consumer behavioral responses | Interactivity, visualization, entertainment, professionalization, Gender of consumer, Platforms | Questionnaire | 454 | SOR + PLS-SEM |
| (Zhang et al., 2020) | Investigate the factors and ways that affect consumers' purchase intention in ELS | Information quality and interaction quality | Questionnaire | 207 | Social exchange theory |
| (Mao, 2022) | Investigate how distinct types of streaming contents influence viewers' psychological and behavioral responses | Type of live broadcast, the consumers' psychological satisfaction and social satisfaction | Questionnaire | 583 | Hypothesis testing |
| (Wu et al., 2023) | Investigate how the improvisation ability can affect the performance of live streamers | Improvisation ability | Video of ELS | 314 | Expert assessment methods + OLS |
| (Y. Wang et al., 2022) | Investigate the influence of product information and live shopping atmosphere on consumer purchase intention | Product information and shopping atmosphere | Questionnaire & interview | 240+16 | Hypothesis testing |
| (Wu & Huang, 2023) | Investigate how to better stimulate consumers' continuous purchase willingness in ELS | Utilitarian value, hedonic value and social value, and consumers' trust | Questionnaire | 213 | SOR |
| (Chen et al., 2023) | Investigate how social presence affects consumers' purchase decisions is limited | Social presence | Questionnaire | 390 | Social presence theory |
| (Ng et al., 2022) | Investigate the development of customer satisfaction and cognitive assimilation through ELS | Perceived serendipity, affective and cognitive perspectives | Questionnaire | 453 | PLS-SEM |
| (B. Wang et al., 2022) | Investigate the difference between the impact of live delivery on hedonic and utilitarian products | Type of products: hedonic or utilitarian | Questionnaire | 162 | Hypothesis testing |
| (Shen et al., 2022) | Investigate the consumer stickiness in live streaming e-commerce | Consumers' subjective feelings | Questionnaire | 262 | Expectation confirma-tion theory + PLS-SEM |
| (Ye et al., 2022) | Investigate the factors that affect the choice of ELS platform by customers | Consumers' subjective feelings, objective situation of the product, external influence | Questionnaire | 443 | Push-Pull-Mooring model + SEM |
| (Wang et al., 2021) | Investigate the influence mechanism of the word-of-mouth reputation of streamers | Credibility, professionalism, interactivity and attractiveness | Questionnaire | 218 | Fuzzy-set qualitative comparative analysis |

The literature on live streaming generally recognizes the importance of expertise, social presence, and attractiveness. However, the focus of previous studies has been limited to certain factors, with few studies directly predict the sales value of e-commerce live streaming. Moreover, the data used in live streaming literature has been limited and monotonous, with most studies relying on survey data. Overall, the limitations of inconsistent theoretical perspective and potentially biased survey data suggests that there is ample room to contribute to the current literature.

Machine learning (ML) has gained popularity in microeconomics due to its advantages in automation, prediction accuracy, and scalability. As a result, there has been growing interest in exploring the potential applications of machine learning in microeconomic research. Table 2 presents some recent examples of how machine learning has been applied in various microeconomic fields.

**Table 2**

Literature review of machine learning (ML) in microeconomic.

| Reference | ML method | Best model | Purpose | Model explanation | Data source |
|---|---|---|---|---|---|
| (Ben Jabeur et al., 2021) | RF, Xgboost, NN, LR, LightGBM, DA, catBoost | RF and LightGBM | Predicting whether oil prices crash or not | SHAP | Yahoo Finance Dataset |
| (Javaid et al., 2022) | DT, RF, ET, NN, LR | RF | Predicting whether a customer buys an item | PI | Online shoppers intention (OSI) dataset |
| (Bussmann et al., 2020) | LR and XGBoost | XGBoost | Predicting the credit risk of small businesses | SHAP | European External Credit Rating Agency (ECAI) |
| (Carta et al., 2021) | DT, RF, GBM, MLP | DT | Predicting the magnitude (high or low) of a company's future stock price from documents (news, etc.) | / | Dow Jones "Data, News, and Analytics" Dataset |
| (Moscato et al., 2021) | LR, RF, MLP | RF | Predicting whether a loan on a P2P platform will be repaid | LIME, ANCHORS, SHAP, BEEF, LORE | Lending Club's dataset |
| (Yin et al., 2022) | LightGBM | LightGBM | Predicting the state of several price movements | PI | WIND |
| (Gramespacher & Posth, 2021) | DT, LR | DT | Predicting risk assessment for credit approval | PI | ECAI |
| (Bastos & Matos, 2022) | FRM,DT,GBM | GBM | Predicting risk assessment for credit approval | SHAP, ALE | Moody's Ultimate Recovery Database |
| (Islam & Amin, 2020) | GRF, GBM | GBM | Predicting products' backorder | PI | Kaggle |
| (Dunstan et al., 2020) | SVM, RF, XGB | RF | Predicting obesity prevalence at the country-level based on national sales of a small subset of food and beverage categories | / | Euromonitor Dataset and estimated data |
| (Ismail et al., 2020) | LR, ANN, SVM, RF | SVM | Predicting direction of stock price movement | / | Kuala Lumpur Stock Exchange (KLSE) |
| (Valencia et al., 2019) | NN, SVM, RF | NN | Predicting the price movement of several cryptocurrencies | / | cryptocompare.com and Twitter |

In summary, machine learning has seen increasing use in microeconomics, with several examples of its application demonstrating its ability to handle large and complex datasets and forecast economic indicators. However, its specific application in the context of online commerce, especially in the field of e-commerce live streaming, is still in its infancy. Although previous studies have highlighted the importance of factors such as trust, rapport, product quality, and diversity in determining the success of live shopping, the lack of rigorous empirical testing and limited sample sizes hinder the ability to make definitive conclusions. Moreover, live broadcasting data is often difficult to obtain due to the proprietary nature of streaming platforms. In current livestreaming industry, streamers usually rely on subjective experience to guide live broadcasts and improve their sales value, since it is difficult to quantitatively predict GMV through live data.

To fill this research gap, this study aims to obtain a model that can achieved the best prediction of GMV from live data, and investigate the determinants of livestreaming business success from a microeconomic perspective. Our study contributes to the literature by shedding light on the features that help explain the variation in sales performance of livestreamers. To the best of our knowledge, this research is the first few attempts to collect detailed micro data that distinguish different sellers and different broadcasting events. It is specifically a longitudinal data that spans 579 days and covers all livestreamers in Shopping Globally sector of Taobao, which is the major platform of livestreaming in China. Consequently, the micro data enables us to firstly use machine learning approaches to predict and explain the performance of each single sellers in each time of broadcasting.

Based on the unique dataset containing information on Chinese live streaming e-commerce, we identify a machine learning model that accurately predicts GMV with an error rate of approximately 7%. This model has significant practical value as it allows for estimation of potential revenue for any given livestreaming event. Moreover, we utilize model interpretation methods to open the black box of artificial intelligence, thus gaining a deeper understanding of the factors that influence profitability in livestreaming e-commerce. Through our analysis, we are able to identify several interesting and previously unknown insights. Our findings provide valuable insights for optimizing the sales performance of livestreamers and vitalizing this new mode of economy.

The upcoming sections will be structured as follows: Section 2 describes our methodology, including the dataset used, the machine learning algorithms employed, and the model interpretation methods selected. Section 3 outlines the experimental setup. Section 4 presents the results of our experiments, including a comparison of results from different machine learning models, as well as our explanations based on feature importance. Finally, Section 5 serves as our conclusion.

# 2. Methodology

## 2.1. Data and variables

This paper examines the livestreaming commerce industry in China, with a focus on the Shopping Globally sector of Taobao[1], the country's leading livestreaming platform. To construct our dataset, we collected information on all livestreamers who broadcasted between April 12, 2020 and November 11, 2021, a total of 579 days. We conducted preliminary screening by excluding broadcasters with an overwhelming majority of zero observations, resulting in a set of 75 streamers. From this set, we constructed a longitudinal dataset comprising 19,175 samples, with each sample representing data from a single broadcasting event and containing information on the respective streamer.

Specifically, we extracted the livestreaming data from a website[2] operated and maintained by Huitun Technology Co., Ltd. This company specializes in collecting and analyzing live streaming data from Taobao and other e-commerce platforms, and this corresponds to data source A in Table 3. We collected data without violating the robots agreement and without affecting the normal operation of the Huitun Technology servers. To collect the data for this paper, we used web spider technology, which is a program that can automatically retrieve information from the Internet according to certain rules during program execution. To obtain the data, we used the Selenium[3] library in Python for web requests and the Beautiful Soup[4] library for data processing. Specifically, we recorded the HTML code of the web page obtained through the Selenium library and then processed the saved HTML code using the Beautiful Soup library to extract the relevant information.

---

[1] The website is available at https://www.taobao.com.

[2] Huitun contains live data from multiple sources and the website is available at https://dy.huitun.com.

[3] Selenium is an umbrella project encapsulating a variety of tools and libraries enabling web browser automation, the source code for Selenium is available at https://github.com/SeleniumHQ/selenium.

[4] Beautiful Soup is a Python library for pulling data out of HTML and XML files, the source code for Beautiful Soup is available at https://github.com/wention/BeautifulSoup4.

**Table 3**

Variables description.

| Variables | Operational definition | Abbreviation in figures | Data sources | Mean | Std | Min | Max |
|---|---|---|---|---|---|---|---|
| Gross merchandise volume | Gross merchandise volume | GMV | A | 741508 | 5300665 | 20 | 361000000 |
| Live counts | Count of lives in that day | Live_Counts | A | 1.76 | 0.93 | 1 | 9 |
| Views | Views of this live (user-specific) | Views | A | 15183.24 | 24585.63 | 4 | 798267 |
| Likes | Likes of this live | Likes | A | 53141.91 | 112198.80 | 6 | 3759455 |
| Comments | Comments of this live | Comments | A | 3547.18 | 3914.48 | 2 | 58389 |
| Page views | Pages view of this live (HTML-request-specific) | Page_Views | A | 55304.30 | 90681.31 | 16 | 2412945 |
| Fan growth | Fans increment of this live | Fan_Growth | A | 151.90 | 285.87 | 0 | 9218 |
| Wisdom | Sense of wisdom | Wisdom | B | 16.74 | 12.10 | 2 | 45 |
| Distance | Sense of distance | Distance | B | 32.27 | 1.59 | 29 | 36 |
| Sense of youth | Sense of youth | Youth | B | 14.80 | 1.46 | 11 | 18 |
| Golden triangle | Degree of facial golden triangle | Golden_Triangle | B | 66.93 | 3.90 | 58.20 | 80.80 |
| Number of pulses | Number of pulses of the voice | Num_Pul | C | 1207.10 | 315.55 | 410 | 2052 |
| Number of period | Number of period of the voice | Num_P | C | 1170.62 | 313.00 | 391 | 2011 |
| Mean period | Mean period of the voice | Mean_P | C | 0.0045 | 0.00090 | 0.003 | 0.0078 |
| Standard deviation of period | Standard deviation of period of the voice | SD_P | C | 0.0013 | 0.00045 | 0.00063 | 0.0027 |
| First bandwidth | First bandwidth of the voice | Bw_1 | C | 235.76 | 429.19 | 8.24 | 3522.10 |
| Second bandwidth | Second bandwidth of the voice | Bw_2 | C | 378.94 | 336.50 | 13.98 | 1822.55 |
| Third bandwidth | Third bandwidth of the voice | Bw_3 | C | 716.54 | 716.11 | 46.80 | 4717.90 |
| Fourth bandwidth | Fourth bandwidth of the voice | Bw_4 | C | 852.78 | 853.92 | 68.58 | 3968.12 |
| Mean intensity | Mean intensity of the voice | Mean_I | C | 71.91 | 3.51 | 58.55 | 77.64 |
| Minimum intensity | Minimum intensity of the voice | Min_I | C | 37.34 | 3.21 | 29.40 | 51.86 |
| Maximum intensity | Maximum intensity of the voice | Max_I | C | 82.14 | 2.74 | 70.72 | 87.18 |
| Service attitude | Service attitude | Service | D | 4.83 | 0.076 | 4.5 | 4.9 |
| Logistics | Logistics evaluated by customers | Logistics | D | 4.83 | 0.076 | 4.5 | 4.9 |
| Activeness | Activeness of live room | Activeness | D | 0.63 | 0.19 | 0.10 | 1.00 |
| Favorite streamer | Counts of "Favorite streamer" chosen by fans | Favorite | D | 3036.55 | 4348.70 | 105 | 36500 |
| Enthusiasm | Enthusiasm | Enthusiasm | E | 86.16 | 8.41 | 60 | 95 |
| Elegance | Elegance | Elegance | E | 82.06 | 10.14 | 60 | 95 |
| Physical appearance | Physical appearance | Appearance | E | 80.32 | 9.01 | 60 | 95 |
| Number of streamer | Number of livestreamers | Streamers | E | 1.46 | 0.63 | 1 | 3 |
| Female proportion | Proportion of female livestreamer in one broadcasting event | Female | E | 0.74 | 0.39 | 0.00 | 1.00 |

To quantify the physical attractiveness of livestreamers, we learn from studies in the cosmetic surgery and beauty industry. The facial scoring system[5] developed by So-Young International are commonly used for Chinese young adults to evaluate their attractiveness (Wang et al., 2020; Wen, 2021). Thus, we follow this pattern to quantify the facial aesthetics of our livestreamers, which constitutes data source B in Table 3. In order to evaluate the voice of streamers, we follow the studies in broadcasting and hosting and utilize the Praat program[6] as a tool for analyzing acoustic features related to anatomical structures (Boersma, 2001; Winn, 2020). This constitutes data source C in Table 3. In addition to these sources, we obtained data on service attitude, logistics, activeness, and favorite streamer from their corresponding shops on the website of Taobao, which is data source D in Table 3. The remaining variables related to the livestreaming style were observed and scored manually. They are marked as data source E in Table 3.

[5] Facial scoring system is available in the website of So-Young International: https://www.soyoung.com/apps.

[6] Praat is a multi-functional linguistics professional software and information of Praat is available at https://www.fon.hum.uva.nl/praat.

The correlation matrix of the numerical variables in our study is depicted in Fig. 1 as a correlogram. The numbers in each grid indicate the correlation coefficient between two variables, and the color of the grid corresponds to the coefficient, with lighter colors indicating higher correlation. Our study aims to generate an accurate prediction of GMV, and the correlation matrix reveals that GMV is strongly correlated with most other variables. Additionally, there are significant pairwise correlations among the variables that denote the popularity of the live broadcast, specifically a correlation of 0.71 between page views and likes, and 0.65 between likes and comments. This could be attributed to the fact that customers who view the page are more likely to give comments and likes. Overall, the variable of interest is highly correlated with the variables representing popularity, which may be crucial in determining the predicted outcome of our machine learning models.

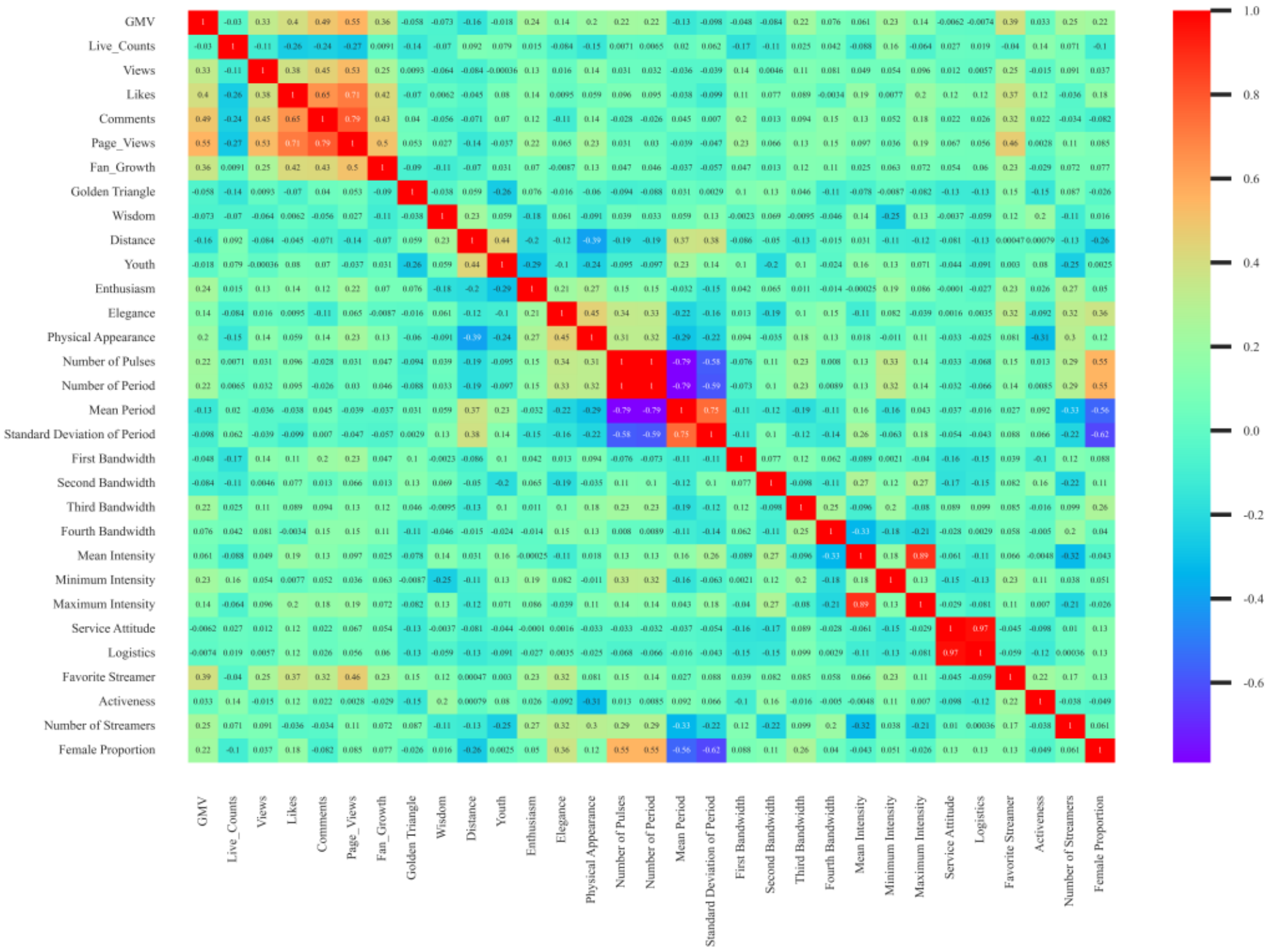

**Fig. 1.** Correlogram of variables.

In order to show the correlation of dozens variable in detail, we divide them into four types, namely, livestreaming popularity, appearance, voice, and miscellany. GMV as target variable is included in all groups. In Fig. 2, four chord diagrams are used to visualize the connectedness. In chord diagrams, the width of a chord indicates the correlation between the two connected variables. And the arc length is determined by the summation of the chord. Fig. 2(a) shows relatively balanced mutual interactions between GMV and variables in popularity category. Also, the importance of “Sense of Youth", "Elegance” in appearance group, “minimum intensity” and “third bandwidth” in sound group as well as “Favorite streamer” in miscellaneous group are shown in the other three diagrams.

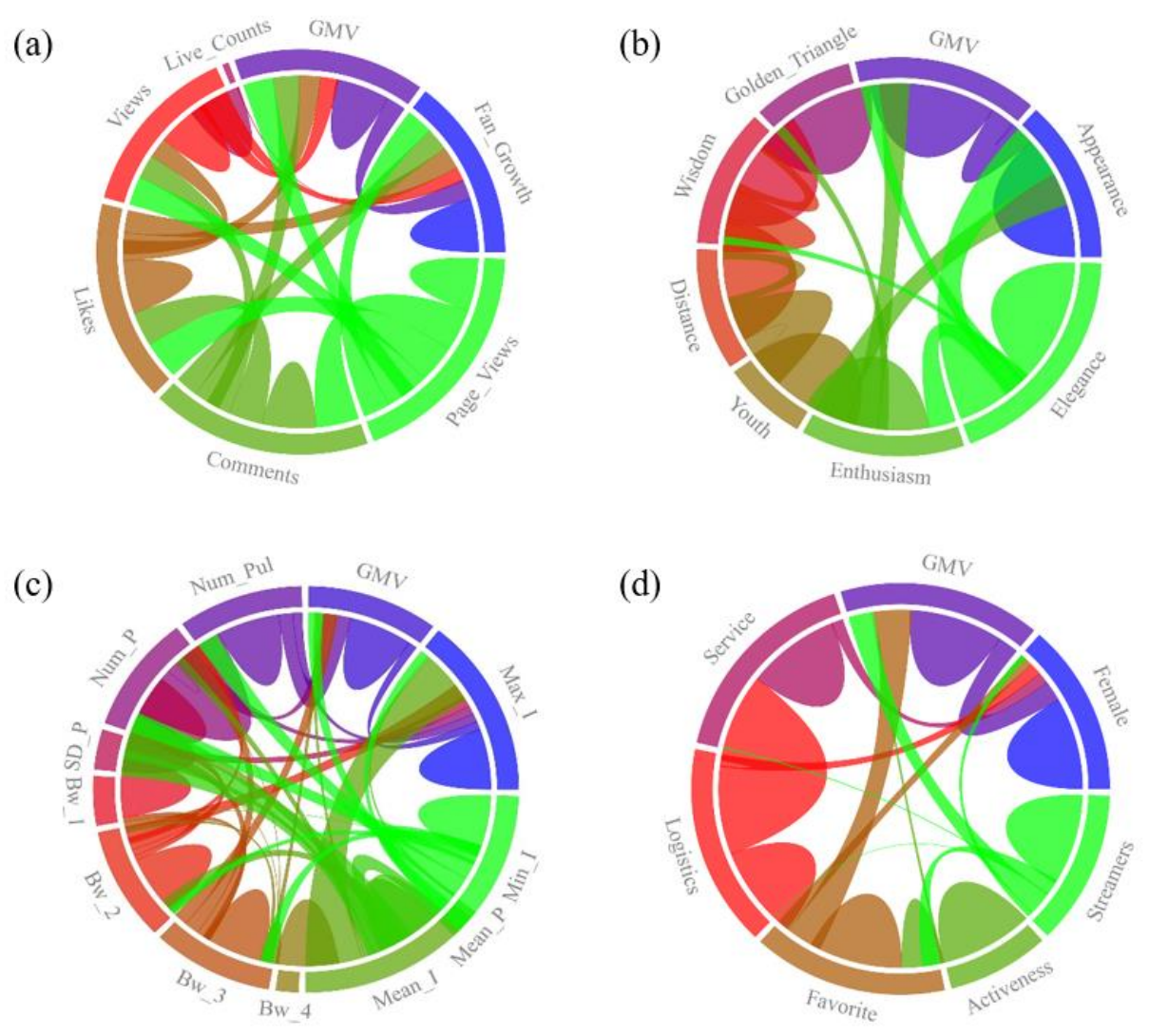

**Fig. 2.** Chord diagrams of variables in livestreaming popularity group (a), appearance group (b), voice group (c), and miscellaneous group (d).

Fig. 3 shows the distribution plot of our target variable, which is natural logarithm of GMV. The brown bars indicate the distribution of target variable, while the blue curve represents a normal distribution with the same mean and variance as target variable. This shows that our data is generally subject to normal distribution.

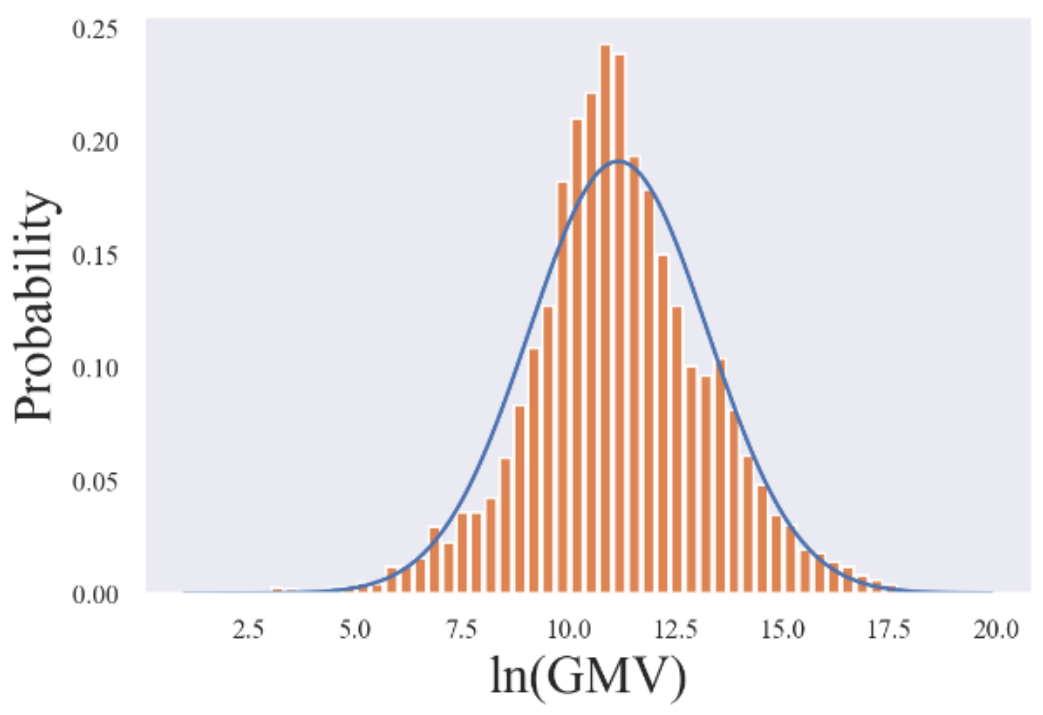

**Fig. 3.** Distribution of target variable.

## 2.2. Machine learning models

In this study, eight machine learning (ML) methods are used to modeling and predicting GMV. For each sample, thirty pre-processed variables of live broadcast data are utilized as input for the ML model, which generates the predicted GMV of the live broadcast as output. ML models learn a mapping between the inputs and outputs of the training set, and the performance of the model is verified on the testing set. We employ 10-fold cross-validation to assess the performance of our machine learning models (Burman, 1989). The best hyperparameters of each ML model are obtained by GridSearch method (Lerman, 1980). Below we give a brief description of all the ML methods used in this research.

## 2.2.1. Decision Tree Regressor

Decision tree (DT) makes prediction by constructing a tree structure and partitioning feature nodes (Safavian & Landgrebe, 1991). In our specific task, where the predicted variable is continuous, the predicted value is computed as the average of all leaf nodes. MSE is employed as the criteria for DT regressor to determine the selecting and splitting features. The minimization problem for choosing the split variable and the split point is as follows:

$$\min_{j,s}\left[\min_{c_1}\sum_{x_i\in R_1\{j,s\}}(y_i-c_1)^2+\min_{c_2}\sum_{x_i\in R_2\{j,s\}}(y_i-c_2)^2\right] \quad (1)$$

Where $j$ is the splitting variable, and $s$ is the splitting point. The smallest combination is achieved by traversing $j$ and scanning $s$ for a fixed $j$.

## 2.2.2. Random Forest Regressor

Random Forest is a popular ensemble learning method based on the bootstrap aggregation approach (Ben Jabeur et al.). In the RF regressor, multiple DT regressors are combined, and there exists no correlation between these individual DT regressors. During the regression task, new input samples are processed by each DT regressor in the forest independently, resulting in a separate regression outcome for each DT regressor. The final prediction of the RF regressor is obtained by taking the mean value of the regression results from all the DT regressors. The performance of the RF regressor is influenced by the hyperparameters of the DT regressors and the number of DT regressors. Increasing the number of DT regressors in an RF model enhances the model's complexity and improves its ability to fit the data.

## 2.2.3. Support Vector Machine

The Support Vector Machine (SVM) model is a feature space-based model that aims to maximize the margin between classes (Hearst et al., 1998). It incorporates kernel tricks, which enhance its nonlinearity as a classifier. The learning strategy of SVM involves maximizing the margin, which can be formulated as a convex quadratic programming problem. This optimization problem is equivalent to minimizing the regularized hinge loss function. In regression tasks, SVM is used as a support vector regressor, which is similar to finding the best-fit line for regression. The learning algorithm of SVM is an optimization algorithm for solving convex quadratic programming problems. The minimization problem of SVM can be defined as:

$$\min_{\beta}\left[C\sum_{i=1}^{N}L_{\epsilon}\big(y_i-f(x_i)\big)+\sum_{j=1}^{p}\beta_j^2\right] \quad (2)$$

Where $L_{\epsilon}$ is an insensitive loss function and $C$ is the penalty assigned to residuals. To solve the minimization problem, a set of weights and a positive definite kernel function that depends on the training data are required. While using radial basis function (RBF) as the kernel of SVM, it is defined as follows:

$$K(x,x')=\exp(-\sigma||x-x'||^2) \quad (3)$$

Where $x$ and $x'$ represent samples, after adding features to get new samples, $K(x,x')$ is to return the calculated value of the new samples.

## 2.2.4. Extra Trees Regressor

The Extra Trees (Margineantu & Dietterich) algorithm bears resemblance to the RF algorithm as both methods employ multiple decision trees (Geurts et al., 2006). However, they differ in terms of their sampling and splitting techniques. While RF utilizes the Bagging model to train individual models and combines their outputs, ET employs all samples and randomly selects features. Furthermore, RF selects the best splitting point from a random subset, whereas ET selects the splitting point completely at random, resulting in a more randomized DT bifurcation process. Consequently, ET demonstrates higher levels of randomness compared to RF and can yield superior results when the dataset is noisy.

## 2.2.5. Logistic Regression

Logistic regression (LR) is a generalized linear model that assumes the dependent variable follows a Bernoulli distribution (Wright, 1995). The LR algorithm first estimates the decision boundary and then establishes the probabilistic relationship between the boundary and regression. In the context of a binary classification task, where we aim to determine the probability y of a given input x being positive, the equation is expressed as follows:

$$P(Y = 1|x) = \frac{1}{1 + e^{-(w^T x + b)}} \tag{4}$$

The logarithmic probability of the output Y=1 is a model expressed as a linear function of the input x. For a regression task, the modified function is as follows:

$$Output = \beta_0 + \beta_1 x_1 + \beta_2 x_2 + \ \dots \ + \ \beta_n x_n + \varepsilon \tag{5}$$

Where $Output$ is the predicted continuous value, $\beta_0$ is the intercept, $\beta_1$ to $\beta_n$ are the coefficients for each binary predictor output $x_1$ to $x_n$, and ε is the error term.

## 2.2.6. K-Nearest Neighbor Regressor

The K-Nearest Neighbor (KNN) algorithm derives predictions by calculating the distance between the sample to be predicted and the known samples, using the distance as a weight and weighting it by the true value of the known sample (Guo et al., 2003). Here shows the Euclidean distance to measure the distance between samples:

$$d(x,y) = \sqrt{(x_1 - y_1)^2 + (x_2 - y_2)^2 + \cdots + (x_n - y_n)^2} = \sqrt{\sum_{i=1}^{n} (x_i - y_i)^2} \tag{6}$$

Where $d(x,y)$ is the calculated distance of sample $x$ and $y$. n is the dimensions of variables. KNN searches for k nearest neighbors in the training set, and the predicted value is the average of the values of the k-nearest neighbors.

## 2.2.7. Adaptive Boosting

Adaptive Boosting (AdaBoost) is an iterative algorithm that generates a series of regressors (Margineantu & Dietterich, 1997). During each iteration, AdaBoost assigns higher weights to regressors with low error rates and lower weights to regressors with high error rates. The output of AdaBoost is calculated based on the weights and outputs of these regressors. In our study, each regressor is configured as a decision tree regressor with a depth of 1. The formula for calculating the output of the AdaBoost

regressor is as follows:

$$output(x) = \sum_{k=1}^{K} \left(\ln \frac{1}{\alpha_k}\right) g(x) \tag{7}$$

Where k is the total number of DT regressors. $\alpha_k$ is the learning weight coefficient. $g(x)$ is the median of weighted outputs of each regressor. The weight update formula for each iteration is as follows:

$$w_{k+1,i} = \frac{w_{ki}}{Z_k} {\alpha_k}^{1-e_{ki}} \tag{8}$$

Where $e_{ki}$ is the error of i-th sample. $Z_k$ is the normalization factor such that the sum of the sample weights is 1.

### 2.2.8. Gradient Boosting Regression Tree

Gradient Boosting Regression Tree (GBRT) is a boosting method that builds upon decision tree regressors (Friedman, 2002). It is an iterative algorithm consisting of multiple decision tree regressors. Each tree makes binary predictions based on a feature of the training data, and subsequent trees aim to minimize the residuals of the predicted values. The model is trained by accumulating the predicted values from all tree nodes, gradually reducing the discrepancy between the predicted and true values. The output of GBRT is obtained through the following additive model:

$$f(x) = \sum_{m=1}^{M} h_m(x; \theta_m) \tag{9}$$

Where M is the number of DT regressors. $\theta_m$ is the parameters of the m-th DT regressor. $h_m(x; \theta_m)$ denotes the contribution of the *m*-th DT regressor to GBRT.

## 2.3. Model Explanation

In addition to obtaining a prediction model with practical significance, we explore the influence of different variables on GMV using model explanation methods. In this paper, Shapley Additive Explanations (SHAP) and Accumulated Local Effects (ALE) are used to interpret the model.

### 2.3.1. Shapley Additive Explanations

SHAP is a model explanation method derived from game theory. Considering a situation where a coalition of players co-create value and reap benefits, SHAP gives a calculation method to distribute the benefits. SHAP allocates expenditure to players according to their contribution to the total expenditure. For a regression model, all input variables contribute to the final prediction, so every variable is a “player” in the coalition. The prediction is the co-create value of coalition. The importance of a variable, known as SHAP values, is measured by how much it contributes to the prediction. For sample $x$, SHAP value of variable $j$ is calculated by following formula:

$$\Phi_j(val) = \sum_{S \subseteq \{x_1,\dots,x_p\}\{x_j\}} \frac{|S|!(p-|S|-1)!}{p!} \left(val(S \cup \{x_j\}) - val(S)\right) \tag{10}$$

Where $\Phi_j(val)$ is the SHAP values of variable j with a specific prediction model $val$. $S$ is a subset of input variables. $|S|$ is the number of variables in subset $S$. $p$ is the total amount of variables in the prediction model. The global SHAP value of variable $j$ is the sum of absolute SHAP values of $j$ among all samples.

### 2.3.2. Accumulated Local Effects

SHAP provides feature importance values for each variable in a given sample, which allows for an understanding of how each variable contributes to the prediction. However, it does not explicitly reveal the direct relationship between input variables and the resulting prediction. ALE shows a visual representation of how a variable influences the prediction of regression model. An ALE plot shows the function between variable of interest and the prediction of regression model while keeping other input variables constant. ALE has two advantages including (1) the interpretation is clear. The horizontal axis of the graph represents the value of predictor and the vertical axis represents the value of predicted outcome, which resembles best fitting line in regression. (2) By firstly differentiation and then accumulation, ALE avoid the making of unlikely data values in PDP and ICE, which are the traditional interpretation methods, while at the same time distinguish the effect of one feature from the joint effects. This suits our data since we show strong correlation between several variables in Fig. 1. For an ALE plot of variable j, we first divide its range using a grid with K bins. $\{Z_k\}_{k=1}^{K}$ denotes the set of values that define the grid. In our paper, $Z_k$ are chosen as the 20-quantiles of the empirical distribution of variable j, where $Z_1$ is the lower bound of variable j, and $Z_{20}$ is the upper bound of variable j. The decentralization effect function is:

$$\hat{f}_{j,ALE}(x) = \sum_{k=1}^{k_j(x)} \frac{1}{n_j(k)} \sum_{i:x_j^{(i)} \in N_j(K)} [f(z_{k,j}, x_{Jay}^{(i)}) - f(z_{k-1,j}, x_{Jay}^{(i)})] \tag{11}$$

Where $\mathrm{n_j(k)}$ is the number of data points in common in the grid, which is 958 in our plots. $Z_{k,j}$ is the value of the upper bound of j, $Z_{k-1,j}$ is the value of the lower bound of j. Further centralizing the effect function, where the average effect is zero, yields the following formula:

$$\hat{f}_{j,ALE}(x) = \hat{f}_{j,ALE}(x) - \frac{1}{n}\sum_{i=1}^{n} \hat{f}_{j,ALE}(x_j^{(i)}) \tag{12}$$

That is, the average effect is subtracted from each effect, so that a positive feature effect means that the feature has a positive effect on the prediction.

# 3. Results

## 3.1. Experiment set

We use eight machine learning models in our prediction to obtain a best forecasting performance. Each regressor has several hyperparameters that need to be set and tuned. To obtain optimal hyperparameters, we use all possible combinations of hyperparameters values to estimate the optimal set of hyperparameters for each regressor. And the optimal result of the regressor is obtained on the best hyperparameters of each regressor and the final result is obtained by ten-fold cross-validation. Variables were transformed by taking the logarithm prior to being input into the model. The parameter ranges for grid search of various machine learning algorithms are set as follows:

**Table 4**

Hyperparameter ranges.

| Model | Hyperparameter | Range |
|---|---|---|
| DT | Max depth | 5 to 100 |
| | Min samples split | 1 to 10 |
| | Min samples leaf | 1 to 4 |
| | Max features | Square root/ Log2 of variables number |
| RF | Estimators numbers | 10 to 500 |
| | Max depth | 5 to 100 |
| | Min samples split | 1 to 10 |
| | Min samples leaf | 1 to 4 |
| | Max features | Square root/ Log2 of variables number |
| | Bootstrap | True/False |
| SVM | Kernel function | Linear/ Polynomial/ RBF/ Sigmoid |
| | C | 0.1 to 1000 |
| | Epsilon | 0.01 to 100 |
| ET | Estimators numbers | 10 to 800 |
| | Max depth | 5 to 100 |
| | Min samples split | 2 to 10 |
| | Min samples leaf | 1 to 4 |
| | Max features | Square root/ Log2 of variables number |
| | Bootstrap | True/ False |
| LR | Keep intercept | True/ False |
| | Normalize | True/ False |
| | Max iterations | 100 to 5000 |
| KNN | Neighbors | 2 to 50 |
| | Weights | Same weight/ Weight is inversely proportional to the distance |
| | Bounded radius nearest | Linear scan/ KD tree/ Ball Tree/ Weighted |
| | Leaf size | 10 to 50 |
| AdaBoost | Estimators numbers | 10 to 800 |
| | Learning rate | 0.01 to 100 |
| | Loss | Linear/ Square/ Exponential |
| GBRT | Estimators numbers | 10 to 800 |
| | Max depth | 5 to 100 |
| | Learning rate | 0.01 to 0.5 |

## 3.2. General performance of models

We evaluated the performance of our proposed model using three common evaluation metrics: mean absolute error (MAE), mean squared error (MSE), and mean absolute percentage error (MAPE), which measure the average absolute difference, average squared difference, and average percentage difference between predicted and actual values, respectively. These metrics are widely recognized as the standard and interpretable tools for evaluating the performance of machine learning methods in the context of regression problems. (Hastie et al., 2009).

$$MAE = \frac{1}{n}\sum_{i=1}^{n} | y_i - \hat{y}_i | \tag{13}$$

$$MSE = \frac{1}{n}\sum_{i=1}^{n} (y_i - \hat{y}_i)^2 \tag{14}$$

$$MAPE = \frac{1}{n}\sum_{i=1}^{n} | \frac{y_i - \hat{y}_i}{y_i} | \times 100\% \tag{15}$$

Where $n$ is the total number of samples, $y_i$ is the true value of the $i$th sample, and $\hat{y}_i$ is the predicted value of the $i$-th sample. MAE measures the absolute difference between the true values and the predicted values, while MSE measures the squared difference. MAPE is expressed as a percentage and represents the average absolute percentage difference between the true and predicted values.

**Table 5**

Performance of ML methods.

| | DT | RF | SVR | ET | LR | KNN | AdaBoost | GBRT |
|---|---|---|---|---|---|---|---|---|
| MAE | 0.8943 | **0.7518** | 0.8949 | 0.8025 | 1.0184 | 0.9325 | 1.1174 | 0.8365 |
| MSE | 1.6393 | **1.2119** | 1.7255 | 1.3126 | 1.9175 | 1.75 | 2.1756 | 1.4082 |
| MAPE | 8.90% | **7.53%** | 9.02% | 8.03% | 10.18% | 9.25% | 11.06% | 8.37% |

Table. 5 compares three types of loss obtained by eight regressors under ten-fold cross-validation. Among them, RF has achieved the best results with 0.752 of MAE, 1.212 of MSE and 7.53% of MAPE. This means for a new observation of live broadcaster, RF can predict its GMV with merely 7% of MAPE.

## 3.3. Interpreting the black-box

We compared the performance of eight machine learning models and choose random forest model as the best model in predicting sales performance of livestreamers. Yet, it is not easy to understand the internal mechanism of random forest model since it has hundreds of parameters. Thus, we adopt the machine learning interpretability methods to open the black box. Specifically, we evaluate the importance of features, which are essentially the inputs embodied as column in the dataset, in forecasting the target variable.

### 3.3.1. General SHAP analysis

In Fig. 4, the SHAP values calculated from the random forest model are displayed. Among four groups displayed in Fig. 2, variables representing the popularity of life broadcast is revealed as the most important features in predicting the total value of merchandise, with the factors of comments, page views, likes, and new fans ranked as the top four features in terms of SHAP values. This is intuitive because live commerce essentially relies on a group of consumers who make parasocial interaction through views, comments, and likes.

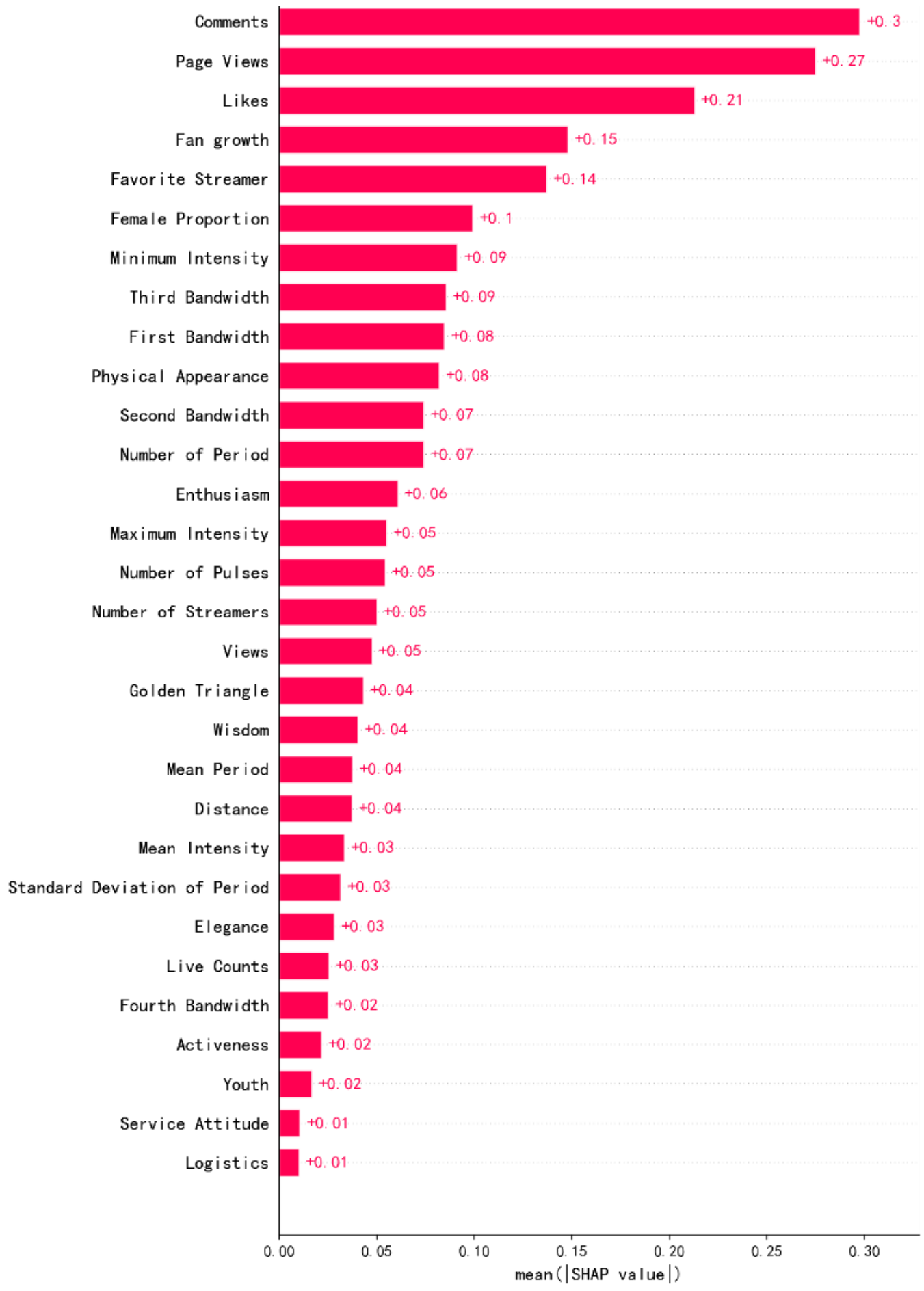

**Fig. 4.** General SHAP values.

In Fig. 4, we also show the influence of variables in auditory and visual groups. Interestingly, the role of voice is even more prominent than that of appearance. Among the sound variables, the information of intensity and bandwidth are closely linked to the sales of streamers, with the minimum intensity of their sound ranked sixth in terms of feature importance. Additionally, the number of pulses and their periods are important characteristics that can refine the prediction precision. This finding is consistent with the general opinion in broadcasting and hosting. Yet our finding extends the impact of the voice from personal attractiveness to market share. In the visual group, physical appearance, enthusiasm, and the degree of facial golden triangle are ranked 10th, 13th, and 18th, respectively, in terms of feature importance. They all contribute to the merchandise volume together with other factors that represent the visual attractiveness of streamers.

### 3.3.2. Gender-specific SHAP analysis

Fig. 5 presents a comparison of the SHAP values between the female-dominated group and male-dominated group. To create these groups, our 19,175 samples were separated based on whether the proportion of female livestreamers was more than 50%. Accordingly, 13,493 samples are included in female-dominated group and 3,495 are in male-dominated group. The remaining 2,187 samples with equal number of female and male livestreamers were excluded from either group. After separating the samples, the RF model was retrained, and SHAP values were calculated for each variable with a 9:1 train-test ratio. In our 1699 samples used for testing, 345 were labeled "male" and 1345 were labeled

"female". We note that Fig. 4 and Fig. 5 demonstrate a strong consensus on the selection of major determinants, while Fig. 5 provides additional insights. For variables such as comments, page views and fan growth, the SHAP values of female-dominated group are significantly higher than its counterpart, indicating that the GMV of female livestreamers depends more on their popularity. On the contrary, for most variables representing voice attractiveness, such as number of period, number of pulses and maximun intensity, the SHAP values of male-dominated group are significantly larger. This finding is interesting since it indicates that the parasocial interaction is more important for female hosts while vocal aesthetics is more decisive for their male counterparts.

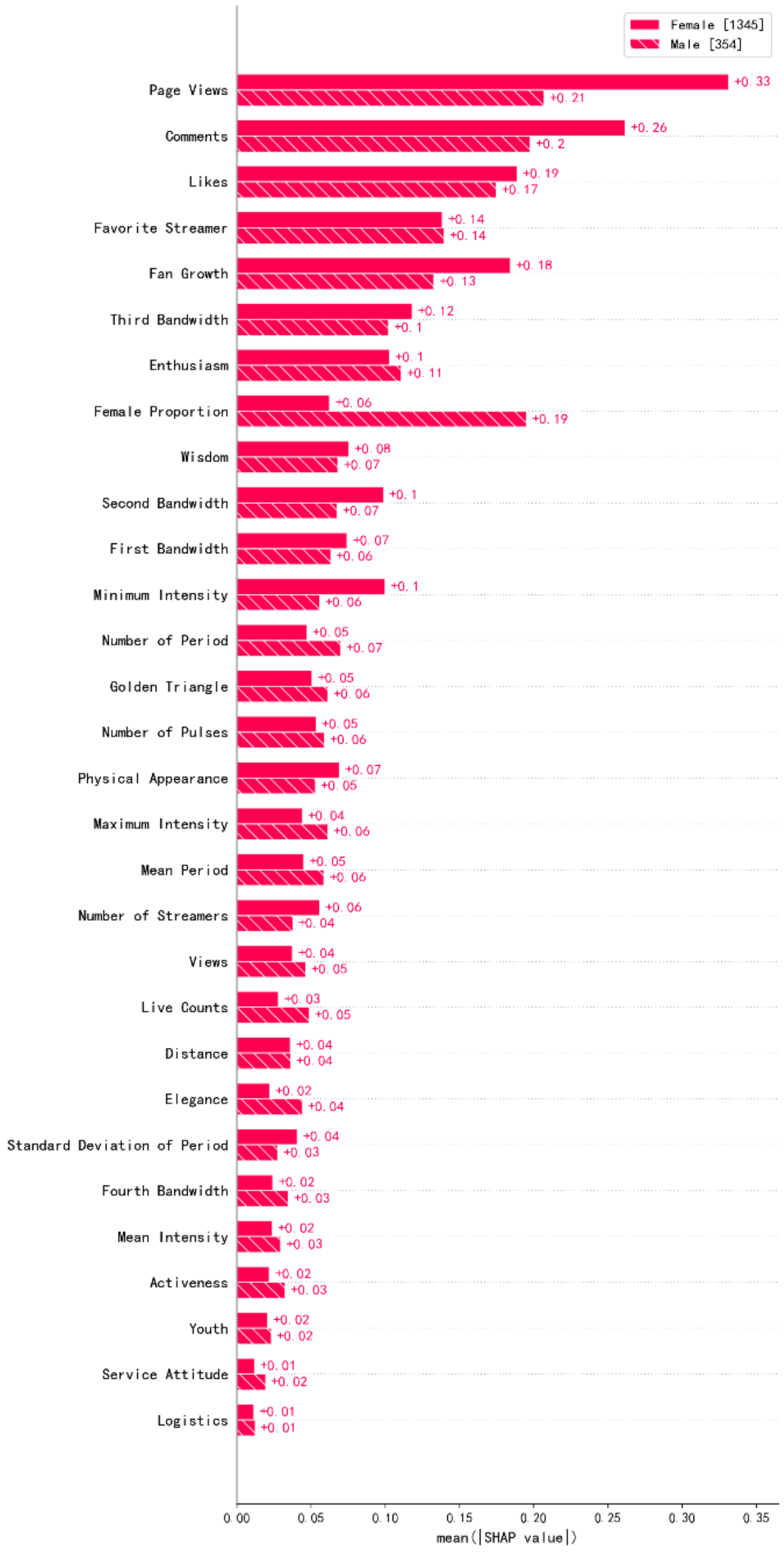

**Fig. 5.** Gender-specific SHAP values.

### 3.3.3. Sample-specific SHAP analysis

Fig. 6 is a SHAP summary plot that shows the feature importance for each variable. The plot displays each observation in the dataset as a dot with two characteristics: (1) the color shows the value of original predictor, with blue for small values, red for large values, (2) the horizontal position of the dot indicating its marginal effect in predicting GMV, with right for positive contribution and left for negative contribution. This approach to calculating contribution was proposed by Shapley (1953) to estimate contribution in n-person games.

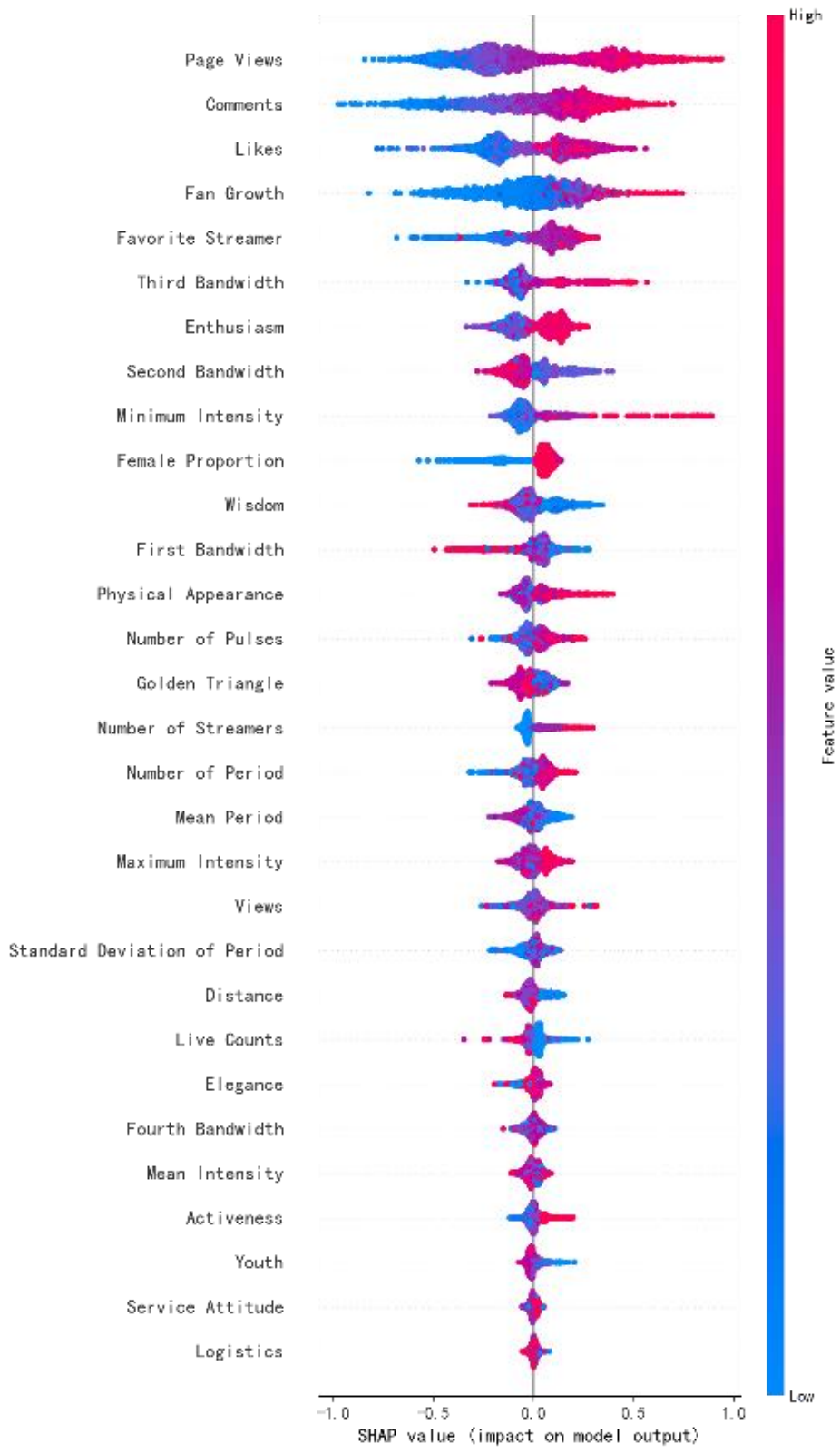

**Fig. 6.** Sample-specific SHAP values.

The results in Fig. 6 are consistent with those in Fig. 4, but Fig. 6 provides sample-specific information. For variables related to popularity, such as comments, page views, likes, and fan growth, the red dots are generally located on the right side, indicating that larger predictors lead to an increased predicted value of GMV, and vice versa. Although the total effects are similar, the distribution of observation shows clear heterogeneity in different variables. For example, the dots of different colors in page views and fan growth are clearly spread out, while the red and purple points in comments and likes are mixed together, showing that the medium and large value of comments and likes may not differ much in predicting the merchandise volume. This suggests that for livestreamers with medium and high level of popularity, real fan growth is more conducive to their sales performance than virtual comments and likes.

For variables related to voice, the SHAP value distributions are largely imbalanced. Dots are clustered on one side and widely spread on the other, indicating that the merit and defect of the voice are not equally valued in the market. This suggests that training in broadcasting and hosting can help increase the sales of streamers. However, the absence of this training is not necessarily an obstacle.

### 3.3.4. Accumulated Local Effects

To show the relationship between GMV and its major influencing factors, we use Accumulated Local Effects (ALE) Plot. ALE Plots describe how the predicted outcome of random forest model fluctuates with changes in the features, namely, page views, comments, and likes. In ALE Plots, The x-axis is predictor while the y-axis is the accumulated change of model prediction. In ALE Plots, the sets of blue lines are accumulated local effect based on each sample, and the black line shows their average effect. In Fig. 7(a) and Fig. 7(b), the average prediction of GMV increases with increasing page views and comments, with relatively flat slopes at both ends. We interpret these periods as bottlenecks for both beginner and top live broadcasters. Fig. 7(c) displays a distinct pattern with an inverted U-shape and blue lines far apart from the average, indicating the complex motivation of people who give likes. For example, some celebrities may receive a lot of "likes" but relatively low merchandise volume, as the audience's interest in watching overshadows their interest in purchasing.

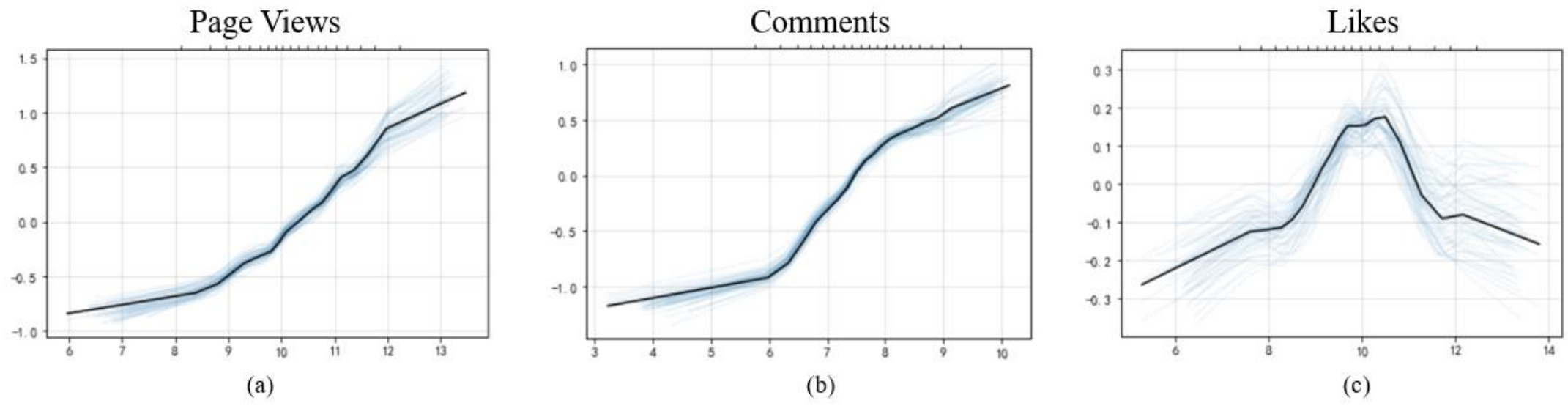

**Fig. 7.** Accumulated Local Effects (ALE) Plots of "Page Views" (a), "Comments" (b), and "Likes" (c).

## 3.3.5. 3D-SHAP analysis

In order to further demonstrate the feature importance of major predictors, we innovatively propose a 3D-SHAP diagram. The x-axis is predictor while the y-axis is the variable of interest, namely, GMV. We add a z-axis that represents the SHAP value of x in predicting y. In this way, we can depict three links including (1) the relationship between a feature and GMV in the plane of classic Cartesian coordinates, (2) the connection between predictor and its feature importance, and (3) the relevance between GMV and the contribution of x in predicting GMV. To avoid the interference of outliers, we use the adjacent-averaging smoothing method with the K nearest neighbor distance as the bandwidth. The smoothed diagrams enable us to describe the fluctuation of variable importance with the change of GMV and its predictors.

We display 3D-SHAP diagrams of three variables that have the greatest impact on GMV, all of which are in popularity group. Fig. 8 shows the 3D-SHAP diagrams of comments, while Fig. 9 and Fig. 10 display the diagrams of page views and likes. We notice that the three figures are generally similar yet with subtle differences. Overall, our three diagrams have similar waterfall shapes, in which slopes are steep in the middle and relatively flat around the top and bottom. This is in line with our findings in SHAP value analyses. However, the shape of 3D-SHAP diagrams of likes is distinctive, in which the top values of likes leads to relatively low SHAP values. This is consistent with our findings in ALE analyses.

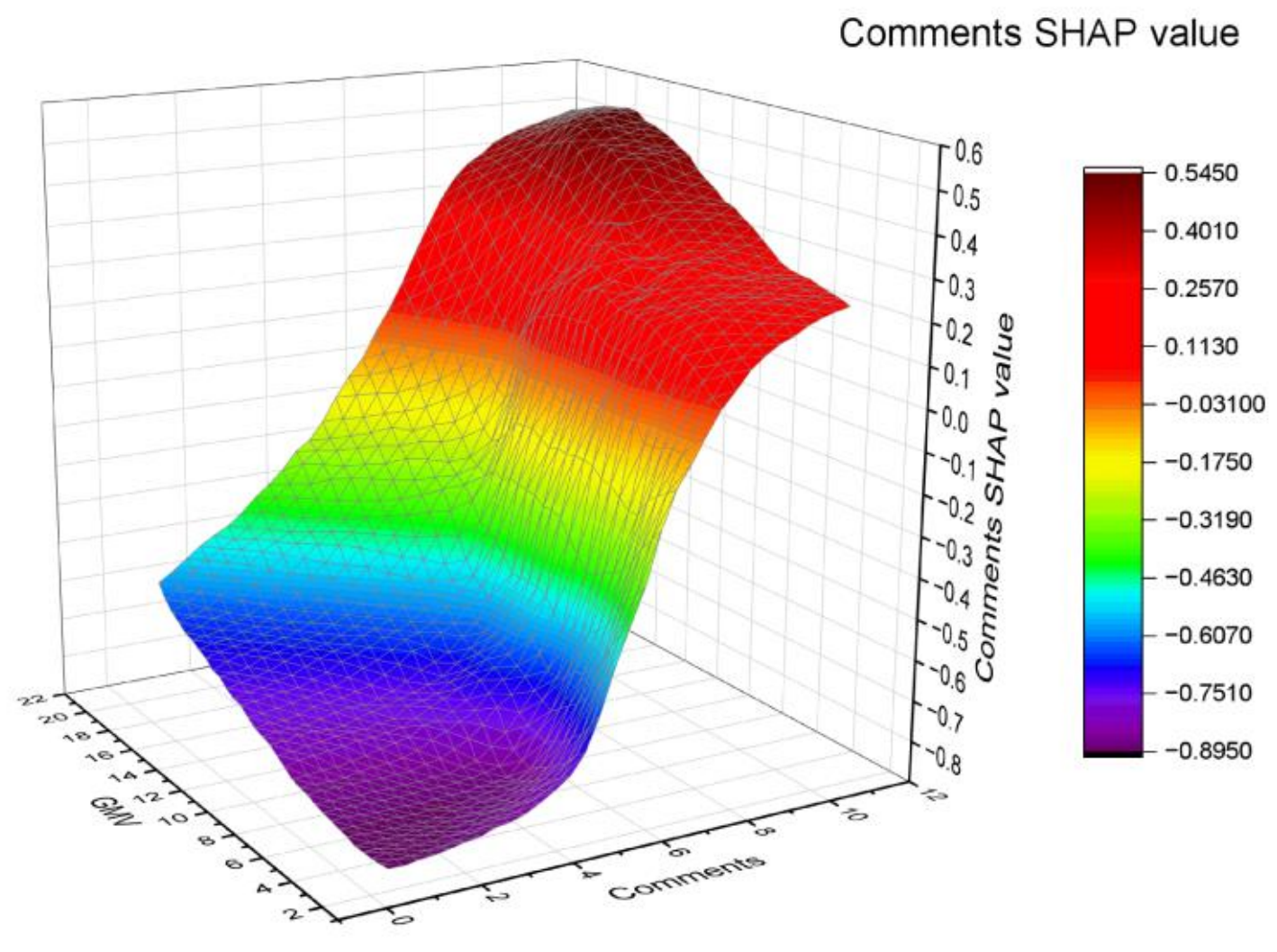

**Fig. 8.** 3D-SHAP of "Comments".

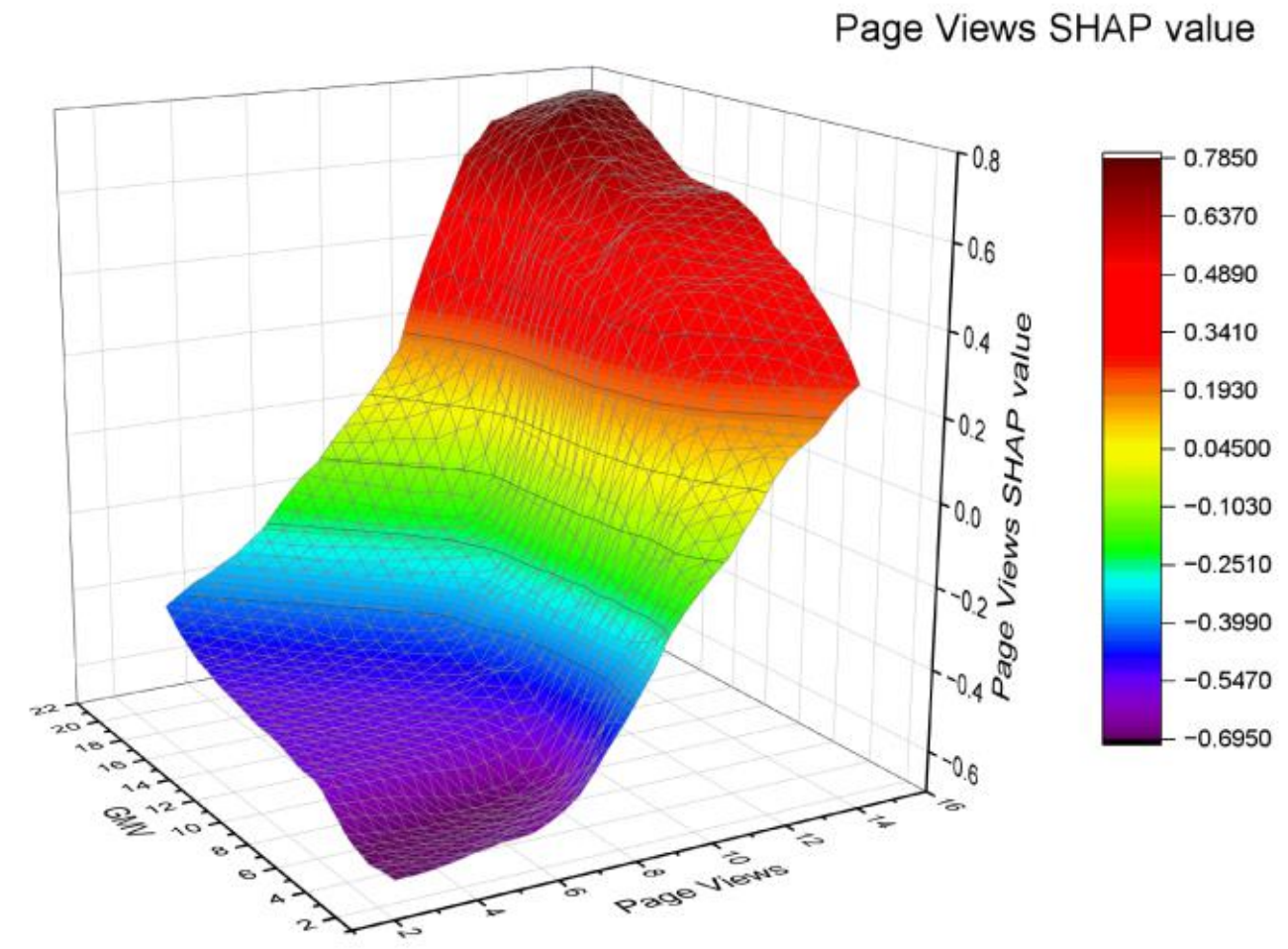

**Fig. 9.** 3D-SHAP of "Page Views".

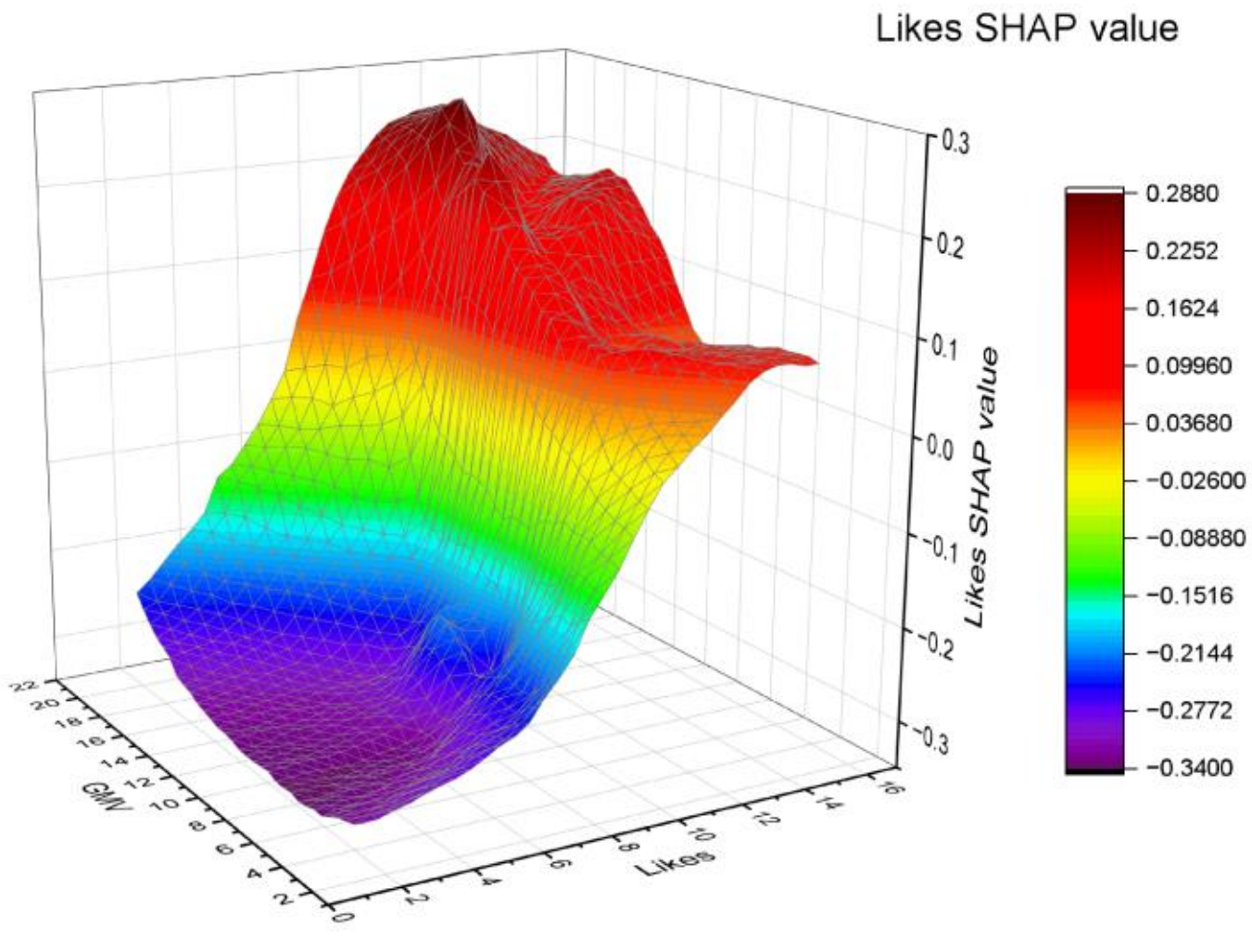

**Fig. 10.** 3D-SHAP of "Likes".

Specifically, Fig. 8 and Fig. 9 are similar in that they all have two inflection points. These two points can be interpreted as two thresholds in the sales of livestreamers, dividing the completely growing period into three stages. The first stage is from the very beginning to the first threshold where log of comments is approximately 4.2 and log of page views is around 7.6, respectively. This indicates that for the samples with popularity lower than the first threshold, they are in the trough time for beginners. Their GMV are not high in this period, and the impact of their accumulated popularity on GMV seemed negligible. It is easy for livestreamers to become disappointed at this stage since their efforts are seemingly futile. However, if their popularity passes the first threshold and come to the second stage, they will witness a fast-growing period, in which the improvement of live popularity is conducive to their GMV expansion significantly. The third stage start from the second threshold, in which log of comments is about 9.5 and log of page views is roughly 12.0. This is a bottleneck period for successful livestreamers, in which the improved popularity is still helpful to their GMV, but the effect is not as good as that in the rising period.

In Fig. 10, we can find similar thresholds around 6.4 and 12.1 for likes. The first threshold of likes assembles those of comments and page views, with its corresponding log of GMV is around 6. Yet the second threshold is unique. We can observe that the second threshold in Fig. 10 comes earlier, where log of GMV of around 14. Additionally, the performance of livestreamers starts to decrease in fluctuations beyond the second threshold, which is shown as a volatile and downward slope in our 3D diagram. This further validates our findings on the effect of likes in ALE, but the 3D-SHAP diagram provides more information. For samples whose likes already entered the third stage, with a GMV less than 14, an increase in likes will shapely decrease their SHAP values. However, if their GMV is greater than 14, an increase in likes will only slightly decrease their SHAP values.

# 4. Conclusion

The fast growth of e-commerce livestreaming has drawn attention from the industry, yet the related research in the scholarly world is still emerging. The main contributions of this study can be summarized as follows. First, we gathered longitudinal data that tracks all non-zero samples of livestreamers. Our data distinguish not only different livestreamers but also different broadcasting events. This data allows us to conducting more in-depth studies including the heterogeneity of livestreamers, and the changes of their features over time.

Second, our research provides an innovative and practical approach in forecasting GMV, which is the major indicator of livestreamers' sales performance. We evaluated the performance of eight machine learning models in terms of their ability to forecast the GMV of any specific broadcasting event. In addition, we find that random forest model can achieve the highest predicting accuracy. This method is meaningful in forecasting the performance of broadcasters and calculating the volume of whole livestreaming sector, thus providing information for the livestreaming platforms and relevant regulatory departments.

Third, we use three different SHAP value analysis and ALE analysis to open the black-box of machine learning model. SHAP value are used to analyze the contribution of each variable to the model prediction and ALE is used to focus on three key variables. In our SHAP value analysis, we find that the top three important features in predicting GMV are variables representing the popularity and variables in auditory and visual groups. Interestingly, the role of voice is even more prominent than that of appearance. When we look into the gender differences, we notice that female livestreamers depends more on popularity, while their male counterpart relies more on voice attractiveness. Additionally, merit and defect of the voice are not equally valued in the livestreaming market. In our ALE analysis that describes how the predicted outcome fluctuates with changes in the features, we find bottlenecks for both beginner and top live broadcasters.

Fourth, we innovatively propose a 3D-SHAP diagram to further demonstrate the feature importance of major predictors. 3D-SHAP diagrams describe the fluctuation of feature importance with the change of target variable and its predictors. We notice that the three figures show similar waterfall shapes yet with subtle differences. In all three graphs, we can see two thresholds in the sales performance of livestreamers, dividing the completely growing period into three stages. However, the second threshold of likes is unique in that the performance of livestreamers starts to fluctuate and decrease after them passing that point. Our findings provide valuable insights for optimizing the commercial value of livestreamers, and could help participants to make full use of the information and opportunities provided by livestreaming platforms.

The framework and database presented above is designed for quantifying the issues that influence the livestreamers' sales performance. One can easily extend that to the intriguing microeconomics topic such as marketing strategy in supplier analysis, shopping experience in demand analysis and relationship theory for both parties. As future work, we plan to refine our quantitative analyze by adopting deep learning based algorithms, thus enriching the literature of retail economy by investigating this emerging mode of commerce.